\documentclass[a4paper,twoside]{article}

\usepackage{epsfig}
\usepackage{hyperref}
\usepackage{subcaption}
\usepackage{calc}
\usepackage{amssymb}
\usepackage{amstext}
\usepackage{amsmath}
\usepackage{amsthm}
\usepackage{multicol}
\usepackage{pslatex}
\usepackage{apalike}
\usepackage{algorithm2e}
\usepackage[bottom]{footmisc}
\usepackage{caption}
\usepackage{tikz}
\usepackage{SCITEPRESS}  
\usepackage{multirow}
\usepackage{tabularx}

\begin{document}

\title{ADMIn: Attacks on Dataset, Model and Input. A Threat Model for AI Based Software}

\author{\authorname{Vimal Kumar, Juliette Mayo and Khadija Bahiss}
\affiliation{School of Computing and Mathematical Sciences, University of Waikato, Hamilton, New Zealand}
\email{vimal.kumar@waikato.ac.nz, jtmm1@students.waikato.ac.nz, khadija.bahiss@waikato.ac.nz}
}

\keywords{Threat Modelling, Artificial Intelligence, Machine Learning, Taxonomy}

\abstract{Machine learning (ML) and artificial intelligence (AI) techniques have now become commonplace in software products and services. When threat modelling a system, it is therefore important that we consider threats unique to ML and AI techniques, in addition to threats to our software.  In this paper, we present a threat model that can be used to systematically uncover threats to AI based software. The threat model consists of two main parts, a model of the software development process for AI based software and an attack taxonomy that has been developed using attacks found in adversarial AI research. We apply the threat model to two real life AI based software and discuss the process and the threats found.}

\onecolumn \maketitle \normalsize \setcounter{footnote}{0} \vfill

\section{\uppercase{Introduction}}
\label{sec:introduction}

While research in Machine Learning (ML) has been actively pursued for several decades, over the last decade or so software products and services that use Machine Learning (ML) and Artificial Intelligence (AI) have gained tremendous visibility. 
ML and AI based software products and services have become ubiquitous and seen in fields as diverse as healthcare, finance, automotive, manufacturing, etc. ML and AI are also extensively being used in cybersecurity for various tasks such as endpoint protection, malware detection, spam filtering, intrusion detection, authentication and fingerprinting etc. In the rest of this paper, we call software products and services that use ML and AI algorithms, AI based software. The use of AI provides advantages of efficiency, functionality and innovation, but it also has the potential to introduce vulnerabilities that are unique to AI based software. The awareness of the risks posed by such vulnerabilities has been steadily growing 
exemplified by the recently held AI safety summit and the Bletchley declaration \footnote{https://www.gov.uk/government/publications/ai-safety-summit-2023-the-bletchley-declaration/the-bletchley-declaration-by-countries-attending-the-ai-safety-summit-1-2-november-2023} at its conclusion as well as president Biden’s executive order on AI\footnote{Executive order 14110: Safe, secure, and trustworthy development and use of artificial intelligence. } .

Practitioners such as security managers, risk management professionals and CISOs also need to be aware of AI risks when they are assessing Information Security risks to their organisations. A crucial aspect of risk management is threat enumeration/identification. To assess AI risks, it is imperative to identify the threats to the AI based software products and services being used in an organisation.

In spite of a large body of literature on adversarial AI and ML, there is a lack of methods or methodologies that practitioners can use to systematically identify threats to AI based software being used in their organisation. Consequently they rely on inconsistent threat identification methods, based either on vendor supplied information or on random threat enumeration.
In this paper we present a methodology to map the existing adversarial AI threats to an AI software development process, thereby creating a threat model that can be used by anyone to detect threats in their AI software. 

The rest of this paper is organised as following. We first discuss work related to threat modelling AI based software in section \ref{rw}. In section \ref{model} we present our software development process diagram for AI based software, while section \ref{section:taxonomy} presents our attack taxonomy. In section \ref{section:tmp} we show how the taxonomy is mapped to the software development process. In section \ref{section:cs} we discuss the case studies of employing our method on real world software before concluding in section \ref{section:conclusion}.



\section{\uppercase{Related work}}
\label{rw}
The purpose of threat modelling is to identify, categorise, and analyse potential vulnerabilities in applications (product, software, data, system, service, etc. ). 
Threat modelling is done by analysing one or more of the attacker’s motivation, goals, techniques, and procedures in the context of the application that is being threat modelled. 
It usually consists of two main parts; modelling and enumeration. In modelling, the entity conducting the exercise creates a model of the system at hand. Common approaches to do this include asset- and impact-centric, attack(er) – and threat-centric, and software- and system-centric approaches. \cite{martins2015towards,selin2019evaluation}. In enumeration, this model is used to identify threats to the system being studied usually aided by a taxonomy. 
 
 There has been substantial amount of work in adversarial machine learning recently, spurred on by the fundamental question on the security of machine learning first asked in \cite{barreno2006can} and \cite{barreno2010security}.  The works described in \cite{worzyk2019physical} \cite{wang2020man} \cite{hu2022membership} \cite{moon2022preemptive} explore techniques that attack specific AI models and their defences, while works such as \cite{wang2022threats}  \cite{greshake2023more} survey and summarise these techniques.  \cite{MIRSKY2023103006} on the other hand conducted a large-scale survey and ranking of threats to AI. Not enough attention, however, has been paid to the development of techniques that can be used to systematically identify threats in AI based software. One of the first attempts to do so was from MITRE who developed MITRE ATLAS (Adversarial Threat Landscape for Artificial-Intelligence Systems) \cite{atlas}. ATLAS is a knowledge-base that classifies adversary tactics and techniques according to the well-known ATT\&CK framework. The European cybersecurity agency ENISA also has overarching guidelines on the risks to AI without specifying the methodology to identify those risks \cite{caroline2021securing}. The work most closely related to ours is STRIDE-AI \cite{s22176662}, which is an asset-centric approach for threat modelling AI based software. It relies on conducting an FMEA (Failure Modes and Effects Analysis) on each individual information asset used in the AI based software development process. The information gained from the FMEA analysis is then mapped to STRIDE to enumerate possible threats. It can be argued that an asset-centric approach while useful for the developors and vendors may not be the best approach for the organisations that are the consumers of AI based software. Microsoft has also utilised the FMEA approach along with bug bar to create guidance on security development cycle for AI based software \cite{msaiml}. This approach is similar to the STRIDE-AI approach and uses the STRIDE categorisation of threats. Finally, OWASP has released the Machine Learning Security Top 10 for 2023 \cite{mltop10}, which lists the top attacks on AI based software. 
 There is some overlap in the threats covered in OWASP's list and our taxonomy which suggests that the list could be integrated with our threat model in future, however that integration is outside the scope of this paper. In addition to the list, OWASP documentation provides guidelines on mitigation as well as metrics on risk factors.

 We proffer that an attack-centric approach (as opposed to an asset-centric approach) to threat modelling provides a more straightforward way to relate the current adversarial AI research to software development, as a taxonomy can be developed from the existing literature. In the rest of this paper we present this attack-centric approach.

\section{\uppercase{Modelling}}
\label{model}
The first step in threat modelling is to create an abstract model of the system under investigation. Existing threat models have used a number of approaches for this. STRIDE for example uses data flow diagrams (DFDs) to create a model and augments those diagrams with trust boundaries. 

\begin{figure*}[t]
  \centering
   \includegraphics[width=\textwidth]{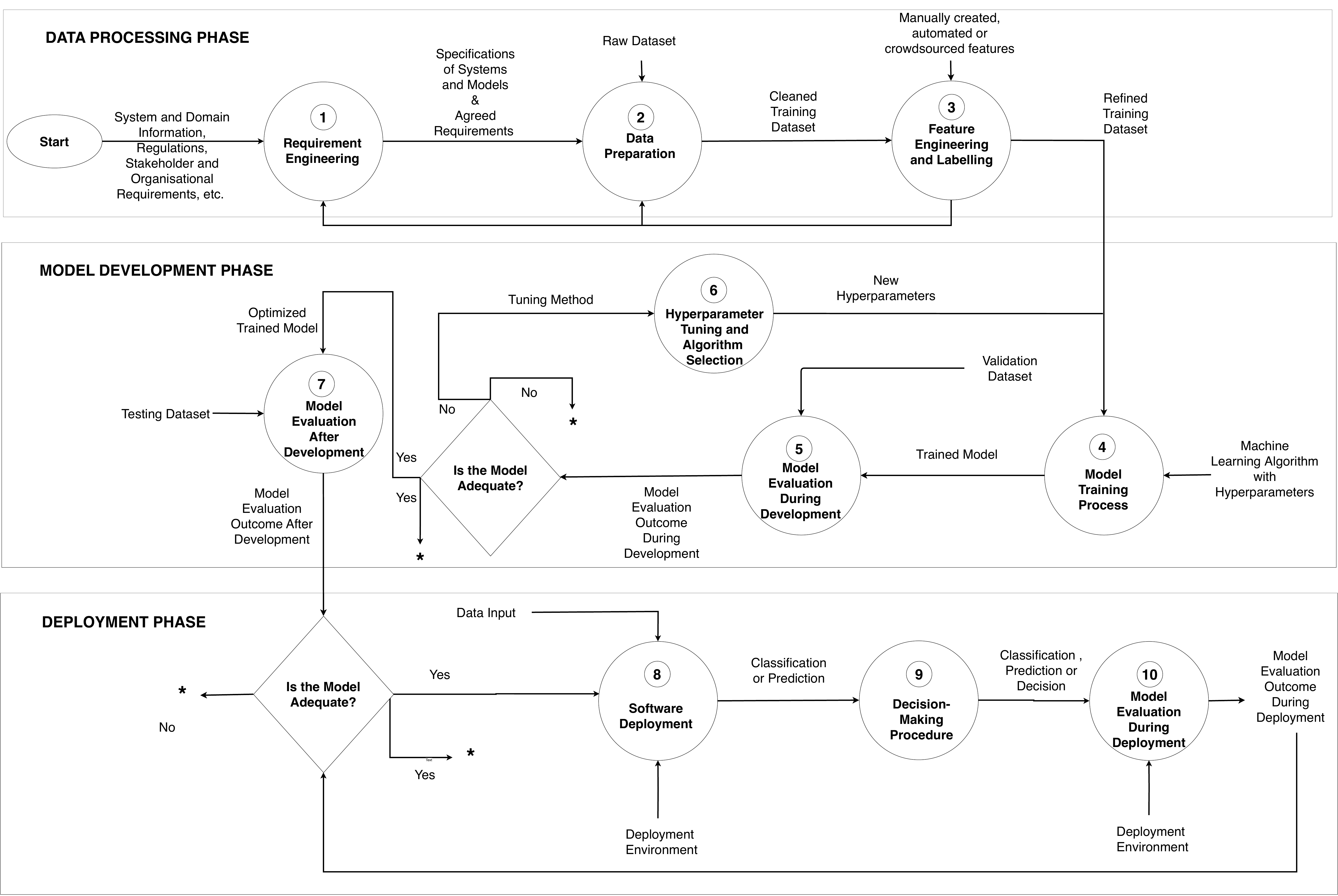}
  \caption{Software development process for AI based software. Circles represent processes, arrows represent inputs and outputs, diamonds represent decisions and `*' means that the arrow can point to any previous process }
  \label{figure:sd}
 \end{figure*}

 We chose to use the software development process of AI based software for modelling. 
 The software development process for AI based software consists of three distinct phases, data processing, model development and deployment, as shown in Figure \ref{figure:sd}. In the diagram presented in this paper we have tried to strike a balance between readability for cybersecurity practitioners, and the detail presented in the diagram. 
 In the rest of this section we go over each of the phases one by one. \\




 \subsection{Data Processing}
 \label{subsection:data processing}
 
 The main objective of this phase is to transform or process datasets into a format that can be used to train a model. 
 We divide the work undertaken in this phase into three processes, Requirement Engineering, Data Cleaning and Feature Engineering \& Labelling.

 \textbf{Requirement Engineering:} The process of requirement engineering involves developers determining the specifications of the client’s AI based software and the requirements of needed datasets. According to \cite{ur2013analysis} there are three inputs into the requirement engineering process: system/domain information, stakeholder/organisational requirements, and regulations. Such requirements are often gathered via several different methods, such as traditional (e.g., interviews), modern (e.g., prototyping), cognitive (e.g., card sorting), group (e.g., brainstorming) or contextual (e.g., ethnography) categories. In general, the outputs of the requirement engineering process are the agreed requirements and the specifications of the systems and model being developed. Other, more specific details may be included as well, such as the plans for acquiring the data, the amount of data needed and how accurate the model needs to be. 

\textbf{Data Preparation:} Once the specification of the software and the requirements of the needed datasets have been identified, work on collecting and cleaning data is usually started. \cite{roh2019survey} have divided the various methods of raw data collection into three categories, discovery, augmentation and generation. The raw dataset thus collected can be in various forms, such as, audio, video, text, time-series or a combination of such formats. It may also have errors, inconsistencies, omissions and duplications. Data cleaning involves finding and resolving these errors, inconsistencies, omissions and duplications. Data cleaning is a fundamental part of this process, and is often used in combination with data collection and curation \cite{symeonidis2022mlops}. Data preparation sometimes involves other techniques such as data transformation and is a vital step in the data processing phase  \cite{kreuzberger2023machine}.

\textbf{Feature Engineering \& Labelling:} Features are elements used to describe specific objects in data. The process of feature engineering involves creating features for a dataset so it can be understood and used by an algorithm \cite{dong2018feature}. The Feature Engineering \& Labelling process in our diagram may additionally encompass related techniques of feature extraction, feature construction, feature storage and feature encoding. There may be algorithms that do not have a feature engineering part to them. Our model, however, is created to be exhaustive so that it covers most possibilities. As will be shown later, when this diagram is used, processes that aren't applicable to a given scenario, can be removed.

Labelling is a related idea, often used in supervised or semi-supervised learning and involves assigning a category or tag to a given piece of data \cite{grimmeisen2022visgil}, to help the model learn or distinguish between objects.

 \subsection{Model Development}
 \label{subsubsection:modelling}

The main objective of this phase 
is to train a model and evaluate its performance.
We divide the work undertaken in this phase into four processes, Model Training, Model Evaluation during Development, Hyperparameter Tuning and Model Evaluation after Development.

\textbf{Model Training:} The refined training dataset, and features or labels produced from the preceding process are used as inputs to the Model Training process where an algorithm is trained on the data provided. Another input to this process is an algorithm or model that is to be trained. Depending on the specific details of the AI model, the used algorithms will differ. Examples of some algorithms that can be used include neural networks, ensemble learning and other supervised, semi-supervised or unsupervised learning methods.
Model training is the most critical  process in the development of AI based software and outputs a trained model to make classifications or predictions.

\textbf{Model Evaluation during Development:} In this process, the trained model from the preceding process is used as an input along with a validation dataset. The validation dataset is used on the trained model to measure model performance. This dataset can be generated via several different methods. One method is to split the training dataset into three subsets: the training dataset, validation dataset and testing dataset. Other methods include \textit{k}-fold cross validation, which involves splitting the dataset into `\textit{k}' subsets. In some cases, multiple methods may be used.

\textbf{Hyperparameter Tuning:} If the outcome of model evaluation during development is not adequate or the developers want to improve model performance, the process of hyperparameter tuning may occur. Some examples of hyperparameters that are tuned are, the learning rate, neural network architecture or the size of the neural network used \cite{feurer2019hyperparameter}. Alternatively, developers may also go back to the data cleaning or the feature engineering \& labelling process or change the algorithm used to create the model. In Figure \ref{figure:sd} this is shown by a `*'. This process occurs iteratively until the model is deemed satisfactory by the developers.

Various different types of tuning methods exist, each with their own advantages and disadvantages. Examples include random search, grid search, or Bayesian optimisation. Meta-heuristic algorithms such as particle swarm optimisation and genetic algorithms are other popular tuning methods used as well \cite{yang2020hyperparameter}. 

\textbf{Model Evaluation after Development:} At this stage the model is evaluated once again. This process takes two inputs, the optimised trained model produced after tuning and a testing dataset. The testing dataset is used to assess the performance of the final optimised trained model. If the outcome of the evaluation is adequate, the deployment phase is executed. Otherwise, depending on the situation, the model may need to be retrained from the very beginning, or use different training data, features, or labels.

 \subsection{Deployment}
 \label{subsubsection:deployment}

In this phase the model is deployed as part of a software product or service in environments such as cloud, server, desktop or mobile device. The work undertaken in this phase is divided into three
processes, Software Deployment, Decision Making and Model Evaluation during Deployment.

\textbf{Software Deployment:} This process involves the fully developed AI based software being deployed in different environments. The input into this phase is the data the software uses. This data is used by the software to output a classification or prediction, depending on the problem that is being solved.

\textbf{Decision Making:} While in some cases the classification or prediction may be the desired end-goal, in other cases the classification or prediction output may be fed into a process, which produces a decision based on the input.

\textbf{Model Evaluation during Deployment:} To ensure that the model does not drift overtime and is fit for purpose, constant, iterative evaluation or monitoring of a model during deployment is sometimes necessary. The Model Evaluation during Deployment process encapsulates this thinking. If the evaluation outcome is adequate, the deployment phase is continued. If the evaluation outcome is not adequate, the model may be retrained from the start, or use different training data, features, or labels. This evaluation is usually done periodically and not necessarily after each run during deployment.

\section{\uppercase{AI threat enumeration}}
\label{section:taxonomy}
\begin{figure*}[t]
  \centering
   {\epsfig{file = 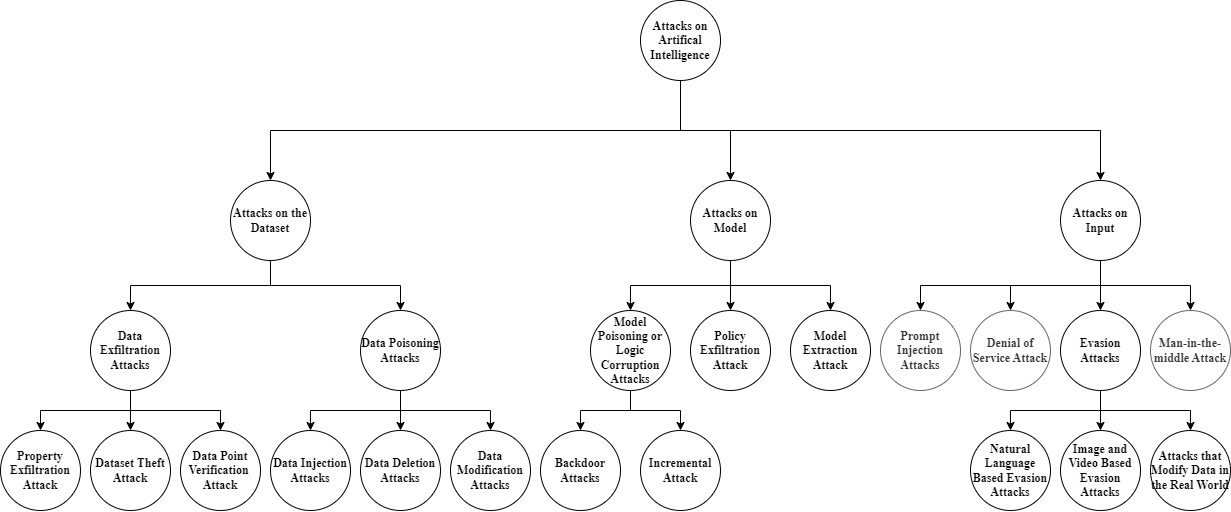, width=\textwidth}}
  \caption{Taxonomy of threats to AI}
  \label{figure:taxonomy}
 \end{figure*}

The second part of threat modelling is threat enumeration. To understand the threats to AI, we explored extensive research literature in adversarial AI. Our literature review has led to the creation of a taxonomy of threats to AI shown in Figure \ref{figure:taxonomy}. In our taxonomy, all possible threats to AI based software are divided into three main categories, attacks on data, attacks on model and attacks on inputs, from which we derive our acronym ADMIn.

\subsection{Attacks on Data}
In these type of attacks, the adversary's focus is on data. The adversary either attempts to steal proprietary data through the algorithm or tries to poison or maliciously modify internal data and/or systems. This category is further split into two types of attacks, data exfiltration attacks and data poisoning attacks.

\textbf{Data Exfiltration Attacks:}
In these attacks, the adversary attempts to steal private information from the target model’s dataset. This can take place in three different ways. First, through property exfiltration attacks, where the attack consists of an adversary stealing data properties from the training dataset. Second, through dataset theft attacks, where the attacks involve the theft of the entire dataset. Finally, exfiltration can be achieved through datapoint verification attacks. In these attacks, an adversary attempts to determine if a specific datapoint is in the model’s training dataset via interactions with the model.

\textbf{Data Poisoning Attacks:}
In these attacks, the adversary deliberately attempts to corrupt the datasets used by the AI based software. The adversary may poison the dataset via adding new data, modifying existing data (e.g., content, labels, or features), or deleting data in the model’s training or validation dataset, with the aim of diminishing the model’s performance. An attack consisting of addition of new datapoints into the training data is performed with the intention of adding biases to the model, so it mis-classifies inputs. \cite{oseni2021security}\cite{liu2022threats}. Poisoning of datasets may take place through the environment or through the inputs to the model. Such attacks may either be targeted or untargeted. In a targeted attack an adversary may attempt to, for example have a malware classification model mis-classify the malware as benign. In an untargeted attack, the adversary on the other hand is looking to make the model mis-classify the malware as anything but the actual classification. Attacks where the adversary is looking to modify or delete existing data, can be comparatively harder to mount as such attacks require the knowledge of and access to the training data. Such access and knowledge however can be gained by exploiting software vulnerabilities in the systems surrounding the dataset.

\subsection{Attacks on Model}
In these type of attacks, the adversary's focus is on the model being used. The adversary either attempts to steal the proprietary model or tries to modify it. This category is further split into three types of attacks, model poisoning or logic corruption attacks, policy exfiltration attacks and model extraction attacks.

\textbf{Model Poisoning or Logic Corruption Attacks:}
In these attacks the adversary attempts to maliciously modify or alter the logic, algorithm, code, gradients, rules or procedures of the software. This can result in reduction of performance and accuracy, as well as causing the model to carry out malicious actions \cite{oseni2021security}\cite{benmalek2022security}\cite{wang2022threats}. Such attacks can be hard to defend against but usually require that the adversary has full access to the algorithms used in the model. This makes these attacks less likely to occur.

\textbf{Policy Exfiltration Attacks:}
In policy exfiltration attacks the adversary attempts to learn the policy that the model is enforcing by repeatedly querying it. The repeated querying may make evident the input/output relationship and may result into the adversary learning the policy or rules being implemented.

\textbf{Model Extraction Attacks:}
Also known as model stealing, the adversary in these types of attacks steals the model to reconstruct or reverse engineer it \cite{hu2022membership}.This is usually done by deciphering information such as parameters or hyperparameters. These attacks require the inputs to the model be known to the adversary whereby unknown parameters can be computed using information from a model’s inputs and its outputs \cite{chakraborty2021survey}.

\subsection{Attacks on Inputs}
In these type of attacks, the adversary uses malicious content as the input into a ML model during deployment. This category is further split into four types of attacks, prompt injection attacks, denial of service attacks, evasion attacks and man-in-the-middle attacks.

\textbf{Prompt Injection  Attacks:}
Prompt injection attacks are a relatively new but well-known type of attack. It consists of an adversary trying to manipulate a (natural language processing) system  via prompts to gain unauthorized privileges, such as bypassing content filters \cite{greshake2023more}. The ChatGPT service for example responds to text prompts and may contain text filters for commercial sensitivity, privacy and other reasons. However, crafting prompts in certain ways may allow users to bypass these filters in what is known as a prompt injection attack. Prompt injection attacks can be harder to defend against compared to other well known injection attacks such as SQL or command injection because the data input as well as the control input, both consist of natural language in textual prompts. 

\textbf{Denial of Service  (DoS) Attacks:}
A DoS attack consists of an adversary disrupting the availability of a model by flooding its inputs with illegitimate requests. DoS attacks are widely understood and in general in such attacks the adversary can flood the model with as many inputs as possible to make the software unavailable to others, however an AI based software can be susceptible to another kind of DoS attack where the model is flooded with deliberately manipulated inputs to cause purposeful mis-classifications or errors \cite{oseni2021security}.

\textbf{Evasion Attacks:}
In these attacks the adversary aims to avoid accurate classification by a model. For example, an adversary may craft spam emails in a certain way to avoid being detected by AI spam filters. Evasion attacks can be undertaken via methods such as changing the model’s policy. The techniques used in evasion attacks are usually specific to the types of inputs the model accepts. We have therefore sub-classified these attacks based on the inputs; Natural Language based attacks, image and video based attacks and attacks that modify data in the real world. A threat modeller can check if any category applies to their specific AI based software by looking at the specific type of input the model expects.

\textbf{Man in the Middle Attacks:}
In these attacks, the input or output of a deployed model is intercepted and/or altered maliciously by an adversary \cite{moon2022preemptive}\cite{wang2020man}. This is usually more likely where an AI based software is being used as a service but is also possible when the software is internally deployed as a product. As an example, an adversary can intercept and alter system and network data being input to a model to produce mis-classification that ultimately results in defence evasion.

\begin{figure}[t]
  \centering
   {\epsfig{file = 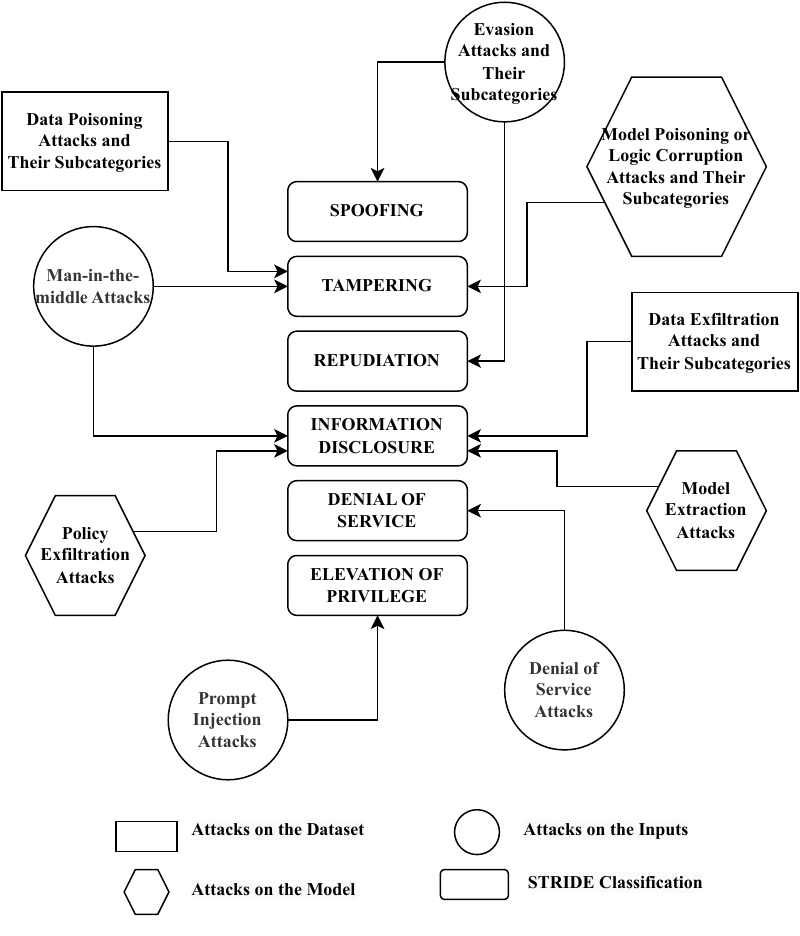, width=\linewidth}}
  \caption{Threats to AI classified according to STRIDE}
  \label{figure:stride}
 \end{figure}

\section{\uppercase{Threats to AI classified according to STRIDE}}
\label{section:stride}
In some cases, one may want to relate each attack or threat to STRIDE if additional information is required. STRIDE is a common threat model used for software threats and as mentioned in section \ref{rw}, threat models from Microsoft and STRIDE-AI use STRIDE. In this section we explain how the threats in our taxonomy are related to STRIDE. This is pictorially shown in Figure \ref{figure:stride}.\\

\noindent\textbf{Spoofing} 
\begin{itemize}
    \item In evasion attacks, the adversary modifies data in the real world or uses other tactics such as modifying images to avoid detection from a model. This is a classic spoofing threat.
    \item In model poisoning, or data poisoning attacks, an adversary may alter code or inputs to a model to avoid detection by creating purposeful mis-classifications.
\end{itemize}
\textbf{Tampering}
\begin{itemize}
    \item In model poisoning, or data poisoning attacks, the main goal of an adversary is to maliciously modify a model's code or data. This is clearly a tampering threat.
    \item In man-in-the-middle attacks, an adversary may tamper with data that is being sent to and from the model.
\end{itemize}
\textbf{Repudiation}
\begin{itemize}
    \item As evasion attacks are considered spoofing attacks, repudiation exists. This is because an adversary can deny carrying out any malicious actions if 
    robust audit logs are missing.
\end{itemize}
\textbf{Information Disclosure}
\begin{itemize}
    \item In data, model or policy exfiltration attacks, the main goal of the adversary is to steal the model, it’s properties or dataset, which discloses confidential information.
\end{itemize}
\textbf{Denial of Service}
\begin{itemize}
    \item In DoS attacks, the adversary’s main goal is to deny the software’s service to users.
\end{itemize}
\textbf{Elevation of Privilege}
\begin{itemize}
    \item In prompt injection attacks, the adversary’s main goal is to gain unauthorized privileges or access via certain input prompts.
\end{itemize}

\begin{table*}[h]
\caption{Comparing the possible threats to the AI based software in the case studies}
\begin{tabularx}{\textwidth}{|l|X|X|}
\hline
\textbf{Attack on} & \textbf{Case study 1}                                                                                                                                                                                                                                                                                                                                                                                                                                                                                                                                                                                                                                                                                                                                                                                                                            & \textbf{Case Study 2}                                                                                                                                                                                                                                                                                                                                                                                               \\ \hline
\textbf{Dataset}   & The application was not susceptible to any of the exfiltration attacks. The software and all its data was open source, aspects such as the algorithm, training data or hyperparameters cannot be stolen as these were already publicly available. 

It was however possible for the dataset to be poisoned. Alhough the dataset is derived from iNaturalist\footnote{\url{https://www.inaturalist.org/}}, which is generally considered a trusted source, it is possible that an adversary can compromise the data repositories to poison the data. 
& The training data for the application was private and stored on cloud. It was discovered that the data was susceptible to exfiltration via dataset theft. The data was collected from customers but it couldn't be said if the data was completely trustworthy. Therefore data poisoning was a valid threat.                                                                                                        \\ \hline
\textbf{Model}     & It was possible for an adversary to poison the model if the adversary had access to some parts of the AI based software development process. Policy Exfiltration and Model Extraction attacks were however not relevant due to the software being open source.                                                                                                                                                                                                                                                                                                                                                                                                                                                                                                                                                           & It  was possible for an adversary to poison the model if the adversary has access to the some parts of the AI based software development process. Policy Exfiltration and Model Extraction attacks were however not relevant due to the software being open source.                                                                                                                                \\ \hline
\textbf{Input}     & As the inputs into this model were images, attacks that involve modifying the model’s environment and image-based evasion attacks could occur. The web-based version of the application was found to be vulnerable to DoS attacks and man-in-the-middle attacks on both the inputs and outputs. The application did not take a prompt as input, therefore, it was not susceptible to prompt injection attacks                                                                                                                                                                                                                                                                                                                                                                                                                                    & As the software is only available to known clients, DoS attacks were not considered to be a threat. Although man-in-the-middle attacks were a possibility, the developers trusted the cloud provider's security enough to not consider it a relevant threat. Prompt injection attacks were not relevant as the model only used images. However, image-based evasion attacks were considered relevant. \\ \hline
\end{tabularx}
\label{tab:cs}
\end{table*}

\section{\uppercase{Threat Modelling Process}}
\label{section:tmp}
AI based software can be threat modelled by mapping the threats in the previously explained taxonomy to the processes in the software development process discussed in section \ref{model}, as described below.

\begin{itemize}
  \item Obtain the software development process diagram for AI based software as shown in Figure \ref{figure:sd} 
  \item Remove any inputs, outputs, or  processes that are/were not used in the software.
  \item Add any additional inputs, outputs, or  processes that are/were used, but are not displayed in the original diagram
  \item Sequentially for each process and its inputs and outputs, add the attacks that apply to your software by referring to the threats in the attack taxonomy shown in \ref{figure:taxonomy}
  \item{If additional information is required, relate the adversarial attacks for each input, process, and output on the ML diagram to STRIDE}
\end{itemize}

\section{\uppercase{Case Studies}}
\label{section:cs}
We used our threat modelling approach on two real world AI based software and went through the process with their developers to evaluate the effectiveness of our approach. In this section, we go over the results of the threat modelling exercises.

\subsection{Case Study 1}
A threat modelling case study was undertaken with the app and website, `Aotearoa Species Classifier' \footnote{\url{https://play.google.com/store/apps/details?id=com.waikatolink.wit_app}}. This software identifies animal and plant species from all around New Zealand from a single photo. We went through the 5 step threat modelling process as described in section \ref{section:tmp} with the developers.

\textbf{Step 1.}We used the software development process diagram from Figure \ref{figure:sd}.

\textbf{Step 2.}The application used a convolutional neural network so the `Feature Engineering \& Labelling' process was removed from the diagram. The `Model Evaluation During Deployment' process was also removed as it was not used.

\textbf{Step 3.} No additional inputs, outputs, processes, or arrows were added to the diagram.

\textbf{Step 4.}The attack taxonomy was applied to the software development process one by one. While the exercise took place by mapping the taxonomy to each process, for brevity we describe the results in Table \ref{tab:cs} in terms of attacks on dataset, model, and inputs.

\textbf{Step 5.}This step was not undertaken as additional STRIDE information was not required.

\subsection{Case Study 2}
A threat modelling case study was undertaken with an image-based AI software. This software is used to improve roads and automatically detect roading problems such as potholes. We went through the 5 step threat modelling process as described in section \ref{section:tmp} with the developers.

\textbf{Step 1.}We used the software development process diagram from Figure \ref{figure:sd}.

\textbf{Step 2.}Since the application development process included data labelling but not feature engineering, inputs related to feature engineering were omitted from the diagram. The process of `Model Evaluation during Deployment' was removed as it was not used. The yes arrow pointing to `*' at the `Is the Model Adequate?' decision was removed, as it was never implemented during the software deployment process.

\textbf{Step 3.}No additional inputs, outputs, processes, or arrows were added to the diagram.

\textbf{Step 4.}The attack taxonomy was applied to the software development process one by one. While the exercise took place by mapping the taxonomy to each process, for brevity we describe the results in Table \ref{tab:cs} in terms of attacks on dataset, model, and inputs.





\textbf{Step 5.}This step was not undertaken as additional STRIDE information was not required.

\section{\uppercase{Conclusion}}
\label{section:conclusion}


A large number of software products and services these days claim to utilize AI, and cybersecurity practitioners are expected to manage cybersecurity risks posed by such software. In this paper we have presented a systematic approach to identify the threats to AI based software. 
Our threat model entitled `ADMIn' is an attack-centric model, that categorises adversarial AI attacks into three categories. These attacks are mapped to the software development process for AI based software, to ascertain the threats that are applicable to the software under investigation. 

Both AI and Cybersecurity are fields that are seeing rapid development and there is increasing awareness of a need for threat modelling AI based software. 
In future, we would investigate the integration of the ADMIn threat model with OWASP ML Top 10 and MITRE ATLAS. We would also like to build upon ADMIn to create a Risk Assessment and Management methodology for AI based software.

\section*{\uppercase{Acknowledgements}}

The authors would like to acknowledge funding from the New Zealand Ministry of Business, Innovation and Employment (MBIE) for project UOWX1911, Artificial Intelligence for Human-Centric Security.

\bibliographystyle{apalike}
{\small
\bibliography{example}}

\end{document}